\documentclass[12pt]{article}
\usepackage{enumitem}
\usepackage[round]{natbib}

\usepackage{mathrsfs}
\usepackage{mathptmx}       
\usepackage{helvet}         
\usepackage{courier}        
\usepackage{graphicx}        
\usepackage{rotating}            
\usepackage{multicol}

\usepackage{amsfonts}
\usepackage{amsmath,amssymb}
\usepackage{bm}
\usepackage{subcaption}
\usepackage{floatrow,fr-subfig}
\usepackage{rotating}
\graphicspath{{Img/}}

\newtheorem{proposition}{Proposition}
\newtheorem{proof}{Proof}
\newtheorem{example}{Example}
\newtheorem{remark}{Remark}
\makeatletter
\renewcommand\section{\@startsection{section}{1}{\z@}%
                                      {-3.5ex \@plus -1ex \@minus -.2ex}%
                                      {2.3ex \@plus.2ex}%
                                      {\normalfont\bfseries}}
\makeatother
\makeatletter
\renewcommand\subsection{\@startsection{subsection}{1}{\z@}%
                                      {-3.5ex \@plus -1ex \@minus -.2ex}%
                                      {2.3ex \@plus.2ex}%
                                      {\noindent\normalfont\emph}}
\makeatother

\begin{document}

\setlength{\baselineskip}{1.5\baselineskip}

\begin{center} \begin{Large} \textbf{Designing experiments for estimating an appropriate outlet size for a silo type problem}\end{Large} \\ [18 pt]
Jesus Lopez-Fidalgo, Caterina May, Jose Antonio Moler
\end{center}

\noindent \textbf{Abstract:} The problem of jam formation during the discharge by gravity of granular material through a two-dimensional silo has a number of practical applications. In many problems  the  estimation of the minimum outlet size which guarantees that the  time to the next jamming event is long enough is crucial. Assuming that the time is modeled by an exponential distribution with two unknown parameters, this goal translates to  the optimal  estimation of a non-linear transformation of the parameters. We obtain $c$-optimum experimental designs with that purpose, applying  the graphic  Elfving method. Since the optimal designs depend on the nominal values of the parameters, a sensitivity study is additionally provided. Finally, a simulation study checks the performance of the approximations made, first with the Fisher Information matrix, then with the linearization of the function to be estimated. The results are useful for experimenting in a laboratory and translating then the results to a larger scenario.  Apart from the application a general methodology is developed in the paper for the problem of precise estimation of a one-dimensional parametric transformation in a non-linear model.
\bigskip

\noindent \textbf{Keywords:} Elfving graphical procedure; Exponential probability model; Fisher Information Matrix; Granular material; Linearization; Non-linear parameter transformation.

\section{Introduction}\label{sect:intro}
Material in granular form  appears in many contexts of applications, as in the pharmaceutical, chemical, food,  agricultural and mining industry (see \citealp{Nedderman}). 
During the discharge by gravity of this material through an outlet, if the size of the outlet is not large enough, the formation of an {\it arch} at some point usually interrupts the flow, causing a {\it jam}. An arch is defined as a structure consisting of particles which are mutually stabilized  (\citealp{Janda}) until an external input of energy breaks their blocking structure and restarts the flow until the next jam happens.

The problem of jam formation during the discharge by gravity of granular material through a two-dimensional silo has been studied in \cite{Janda}, \cite{Amo1} and \cite{Amo2}. In particular, they focus on studying the waiting time that passes between two jamming events, which depends on the outlet size, according to some model. This waiting time is also related with the {\it avalanche}, that is the amount of material dropped between two jamming events. In \cite{Amo1} and \cite{Amo2} the optimal experimental designs to estimate the unknown parameters and to discriminate between models are obtained.

There is a common interest in avoiding a jam at least during a specific period of time. In fact, the event of breaking the arches may be dangerous, expensive or just no affordable. 
Hence, the goal of this paper is the precise estimation of the minimum outlet size  necessary to guarantee that the expected time between two jamming events will exceed a fixed time of interest. Assuming an exponential model as in \cite{Amo1}, this goal determines the problem of finding an optimal design to estimate a non-linear transformation of the unknown parameters.

When the inferential goal is the estimation of a linear combination of the unknown parameters, a $c$-optimal design minimizes the variance of the maximum likelihood estimator (classical references on optimal designs are, for instance, \cite{Atk:2007} and \cite{Pukelsheim}). \cite{Elfv:1952} provided a graphical method to determine $c$-optimal designs of a linear model on a compact  experimental domain, based on the construction of a {\it convex hull}. This method is not easy to use for more than two  parameters, but \cite{Fidalgo2004} provided an iterative procedure  for more than two parameters based on the graphical Elfving technique. For instance \cite{Fidalgo2004} used successfully this procedure to compute c--optimal designs for more than two parameters.
Since the model  considered  is non-linear  it is possible to determine a local $c$-optimal design by considering the Fisher Information Matrix (FIM) for nonlinear models and  a first--order linearization of the function of the parameters to be estimated around some nominal values of the parameters. Moreover, we adopt $c$-optimality for estimating a non-linear transformation of the parameters by linearizing also this function.

The paper is organized as follows. Section 2 presents the problem, sets the  basic notation and explains our method in its generality. Section 3 contains the results on $c$-optimal designs. Section 4 provides a sensitivity analysis. Section 5 contains a simulation study to check the validity of the approximations applied to obtain the results. Section 6 concludes the paper. All the computations have been done with Python 3.7. Codes are provided as supplementary material.

\section{Problem and general method}\label{sect:2}
Consider the problem of falling of particles through a two-dimensional silo presented in  \cite{Amo2} and  \cite{Amo1} and introduced in Section 1. Denoting by $T$ the time between two jamming events and by $\phi$  the size of the outlet at the bottom of the silo, let
\begin{equation}
\label{eq:mean:general}
E[T|\phi]=\eta({\phi; \boldsymbol{\theta}})
\end{equation}
be the mean time between two jamming events, where $\boldsymbol{\theta}$ represents the unknown model parameter. Following \cite{Amo2} and \cite{Amo1}, it is realistic to consider that $T$ has an exponential distribution.
 In particular, given the outlet size $\phi$ of the silo, which is a controlled variable, in the next Section we will assume for the mean function \eqref{eq:mean:general} the model in \cite[eq. (3)]{Amo2}.

 There is a common interest in avoiding a jam at least in a period of time.  Hence, our goal is the precise estimation of the minimum outlet size  necessary to guarantee that the expected time between two jamming events will be greater than a fixed time $T_0$ of interest:
\begin{eqnarray}
E[T|\phi]\ge T_0\label{goal}.
\end{eqnarray}
{\noindent If $\eta(\cdot,\cdot)$ is an invertible function, \eqref{goal} becomes
	\begin{equation}\label{g}
	\phi\ge g(T_0,\boldsymbol{\theta}),
	\end{equation}  
for some inverse function $g$. Thus, we are  interested  in   estimating  $g(T_0,\boldsymbol{\theta})$ which is a non-linear  function of the unknown model parameter. Since $T_0$ is a fixed constant from now on we denote it simply by $g(\boldsymbol{\theta})$.
 
To this aim, assume that an experimenter can observe uncorrelated observations from $n$ experiments,
\begin{equation}\label{eq:model}
t_i=\eta({\phi_i; \boldsymbol{\theta}})+\varepsilon_i,\quad i=1, ..., n.
\end{equation}
 Since $\phi$ is a controlled variable, the $n$ experimental conditions $\phi_1, ..., \phi_n$ can be chosen according to a design $\xi$, that is, a probability distribution on a domain ${\cal X}=[a,b]$:
 \begin{equation*}
 \xi=\left\{\!\!\!\begin{array}{ccc}
 \phi_1 &\cdots & \phi_r \\
 p_1 & \cdots & p_r
 \end{array}
 \!\!\!  \right\},
 \end{equation*}
 with $r \leq n$.
 
 Observe that the model herein considered is non-linear and the errors $\varepsilon_i$ have non-constant variance:
 \begin{equation*}
 Var(\varepsilon_i)=\eta({\phi_i; \boldsymbol{\theta}})^2
 \end{equation*}

  The FIM is defined by
\begin{eqnarray}
M(\xi, \boldsymbol{\theta})=\int_{\cal X} I (\phi, \boldsymbol{\theta})d\xi(\phi),\nonumber
\end{eqnarray}
{\noindent  where
\begin{eqnarray}
I (\phi, \boldsymbol{\theta}) &=& \nonumber -\mathbf{\textrm{E}}_Y\left[\dfrac{\partial^2}{\partial \boldsymbol{\theta}^2 }{\cal L}(\boldsymbol{\theta}; t, \phi)\right]
\end{eqnarray}
is a two by two matrix and $\cal L$  is the log-likelihood function. Since, for an exponential model with mean \eqref{eq:mean:general}, we have
\begin{eqnarray} {\cal L}(\boldsymbol{\theta}; t, \phi)=\log\left(\dfrac{1}{\eta(\phi,\boldsymbol{\theta})} \exp -\dfrac{t}{\eta(\phi, \boldsymbol{\theta})}\right),
\end{eqnarray}
 it follows that the FIM of model (\ref{eq:model})  at one point $\phi$ is
\begin{equation}
I (\phi, \boldsymbol{\theta}) =\dfrac{1}{\eta^2(\phi,\boldsymbol{\theta})}
\nabla \eta(\phi,\boldsymbol{\theta})\,\nabla \eta(\phi,\boldsymbol{\theta})^T,\nonumber
\end{equation}
{\noindent and}
\begin{equation}\label{eq:M}
M(\xi, \boldsymbol{\theta})=\int_{\cal X} \dfrac{1}{\eta^2(\phi,\boldsymbol{\theta})}
\nabla \eta(\phi,\boldsymbol{\theta})\,\nabla \eta(\phi,\boldsymbol{\theta})^Td\xi(\phi),
\end{equation}
 where  the transpose is indicated with the  superscript $T$ and $\nabla $ stands for the gradient.

Equation \eqref{eq:M} is also the FIM of the following linear gaussian and homoschedastic model
\begin{equation}\label{linearization}
t_i= {\boldsymbol{\theta}}^T f(\phi_i; {\boldsymbol{\theta}}^T)+ \epsilon_i,
\end{equation}
{\noindent with}
\begin{equation}\label{eq:f}
f(\phi;{\boldsymbol{\theta}})=\dfrac{1}{\eta(\phi,\boldsymbol{\theta})}
\nabla \eta(\phi,\boldsymbol{\theta}),
\end{equation}
 
  Our goal is therefore to find an optimal design for  precise estimation of  $g(\boldsymbol{\theta})$, that is, a design minimizing the variance of the maximum likelihood estimator (MLE) of $g(\boldsymbol{\theta})$.  

 When the inferential goal of an experiment is an efficient estimation of a vector of unknown parameters $\boldsymbol{\theta}$, an optimal design maximizes a suitable functional of the FIM, $M(\xi, \boldsymbol{\theta})$, because
 its inverse is asymptotically proportional to the  covariance matrix of the MLE ${\boldsymbol{\hat \theta}}$, 
 which is asymptotically unbiased. Some classical references on optimal designs are \cite{Fedo:Theo:1972}, \cite{Paz:86} and \cite{Atk:2007}. An optimal design depends on the value of the unknown parameters except in the case of linear models.

 As mentioned above a $c$-optimal design  $\xi^*_c$,  minimizes the asymptotic variance of a linear transformation $\mathbf c^T \boldsymbol{\theta}$ of the unknown model parameters: 
 \begin{equation}\label{c-opt}
 \xi^*_c=\arg \min_\xi \mathbf c^T M(\xi;\boldsymbol{\theta})^{-1} \mathbf c
 \end{equation}

 A very nice way to compute $c$-optimal designs, especially in two dimensions,  is the geometric Elfving procedure (see \citealp{Elfv:1952}). Such procedure is constructed for estimating a linear transformation $\mathbf c^T \boldsymbol{\theta}$ of the parameters given a linear homoschedastic model $T= \boldsymbol{\theta}^Tf(\phi) + \epsilon$.

Remembering that the MLE estimator of $g(\boldsymbol{\theta})$ is  $g(\hat{\boldsymbol{\theta}})$, 
 let us approximate the non-linear function $g(\boldsymbol{\cdot})$ using Taylor expansion around the true value $\boldsymbol{\theta}_{t}$, so that  we can approximate $g(\hat{\boldsymbol{\theta}})$ with $g({\boldsymbol{\theta}}_t)+\nabla g({\boldsymbol{\theta}}_t)(\hat{\boldsymbol{\theta}}-{\boldsymbol{\theta}}_t)$. The variance of $g(\hat{\boldsymbol{\theta}})$ can be then approximated by

\begin{equation}\label{c-var}
\nabla g(\boldsymbol{\theta}_t)^T\,  M(\xi, \boldsymbol{\theta}_t)^{-1}\,\nabla g(\boldsymbol{\theta}_t)
\end{equation}
and a $c$-optimal design for model \eqref{eq:model}  is a design satisfying \eqref{c-opt} with $\mathbf{c}=\mathbf{c}(\boldsymbol{\theta})$
 given by 
\begin{equation}\label{c:def}
\mathbf c(\boldsymbol{\theta})=\nabla g(\boldsymbol{\theta})
\end{equation}

Notice that two procedures of approximation by linearization have been adopted, 
and that the $c$-optimum design satisfying \eqref{c-opt}
depends on the unknown parameters both through the vector $\mathbf c$ and the FIM, $M(\xi, \boldsymbol{\theta})$. Hence, a nominal value $\boldsymbol{\theta}_0$ guessing the true value $\boldsymbol{\theta}_t$ has to be chosen and the design obtained will be locally optimum. Starting from the design space considered in \cite{Janda} and the values obtained in \cite{Amo2}, the procedure to obtain $c$-optimal designs is developed in detail in the next section.

\section{$c$-optimal designs}

Assume that the time $T$ between two jamming events is exponentially distributed with mean
\begin{eqnarray}\label{model1}
\eta({\phi; \boldsymbol{\theta}})=\dfrac{1 }{C}\exp(L\,\phi^2)-1,\; \phi \in {\cal X}=[a,b],
\end{eqnarray}
{\noindent where $\boldsymbol{\theta}^T=(C,\,L)$, as in  \cite[eq. (3)]{Amo2}.

Our main goal is the efficient estimation of the minimal diameter $\phi \in {\cal X}$ for which (\ref{goal}) holds. If the mean of $T$ is given by \eqref{model1}, this means
\begin{eqnarray}
\eta({\phi; \boldsymbol{\theta}})=\dfrac{1 }{C}\exp(L\,\phi^2)-1\geq T_0,
\end{eqnarray}
that is
\begin{eqnarray}
\phi \geq g(\boldsymbol{\theta})=\displaystyle\sqrt{\dfrac{\log(C(T_0+1))}{L}}\,. \label{gmodel1}
\end{eqnarray}

The extremes of the experimental domain ${\cal X}=[a,b]$ have to satisfy $d<a<b<\phi_C$, where $d$ is the diameter of the granular material and $\phi_C$ is a nominal diameter above which jamming is practically impossible. In theory there is not such a value since there is always a chance of forming an arch, no matter how wide the outlet is. But a practical limit can be assumed and even that value is a parameter of some models  (see \citealp{Amo2} and \citealp{Amo1}).  Moreover, $L>0$ since the time, and therefore $\eta$, between jams is increasing with respect to $\phi$; and $0<C<\exp(L\,\phi^2)$ for any $\phi$  since $\eta$  must be positive. This does not mean constrained estimation, but just practical limits. The data will be in charge of dealing with them. 
Following the method presented in Section 2, we are obtaining here $c$-optimal designs for estimating the bound $g(\boldsymbol{\theta})$ given in \eqref{gmodel1}.
\medskip

\begin{remark}
	An alternative goal could be the estimation of the minimal diameter such that, for  a given value $\alpha$,  $P(T>T_1)\ge 1-\alpha$. However, for an exponential model, this is equivalent to consider (\ref{gmodel1}) with $T_0=-T_1/log(1-\alpha)$ since
\begin{eqnarray}
1-\alpha &\leq& P(T>T_1)=\exp (-T_1 /\eta({\phi; \boldsymbol{\theta}}))
\end{eqnarray}	
and then
\begin{eqnarray}
\eta({\phi; \boldsymbol{\theta}})&\geq & \frac{-T_1}{\log(1-\alpha)}.
\end{eqnarray}	
For instance, for $\alpha=0.05$, $T_0=19.5T_1$. Thus, the problem is equivalent in terms of estimation and designing and this is the relationship between both thresholds. In this particular case the threshold for the probability is about 20 times the one for the expectation.
\end{remark}

The information matrix at a point $\phi$ for model \eqref{model1} is
\begin{equation}
I(\phi,\,\boldsymbol{\theta})=\dfrac{e^{2\phi^2L}}{C(e^{\phi^2\,L}-C)^2}\left(
                                                                     \begin{array}{cc}
                                                                       \dfrac{1}{C} & -\phi^2 \\
                                                                       -\phi^2& C\phi^4 \\
                                                                     \end{array}
                                                                   \right).\label{minf}\end{equation}
{\noindent In order to apply the Elfving's graphical method, we need to obtain the Elfving locus, that is, the convex hull of the union of the curve defined by the regressors in \eqref{linearization} and its reflection through the origin; the $c$-optimal design is then determined by the crossing point between the line $\mathbf c$ and the boundary of the Elfving locus (see \citealp{Fidalgo2004}).  The parametric equations
	that represent the curve are obtained from (\ref{eq:f}), which becomes, when model \eqref{model1} is assumed, $f(\phi,\boldsymbol{\theta})=G(\phi,\,\boldsymbol{\theta})\,(1/C,-\phi^2)^T$, where
	\begin{equation}\label{G}
	G(\phi,\,\boldsymbol{\theta})=\dfrac{e^{\phi^2L}}{e^{\phi^2L}-C}.
	\end{equation}
	 Hence, the parametric equations of $f([a,b])$ are
\begin{equation}\label{param2}
\begin{cases}
x(\phi)=G(\phi,\,\boldsymbol{\theta})/C,\\
y(\phi)=-G(\phi,\,\boldsymbol{\theta})\,\phi^2,\\
\phi\in[a,b].
\end{cases}
\end{equation}	

According to the experimental case considered in \cite{Janda},  $\phi\in{\cal X}=[1.53,\,5.63]$ and   the estimates obtained in \cite{Amo2} from data will be used as nominal values of the parameters, that is, $C_0=0.671741$ and $L_0=0.373098$.
Figure \ref{convex} represents the parametric curve \eqref{param2}, its reflection, and the Elfving locus $\overline{A_1A_2A_3A_4}$ obtained in this case. It is worth observing that the vertexes of the convex hull  in Figure \ref{convex} are not tangential points of the curve but  outermost points of the curve.}
\begin{figure}[h]
	\centering
	\includegraphics[width=8cm]{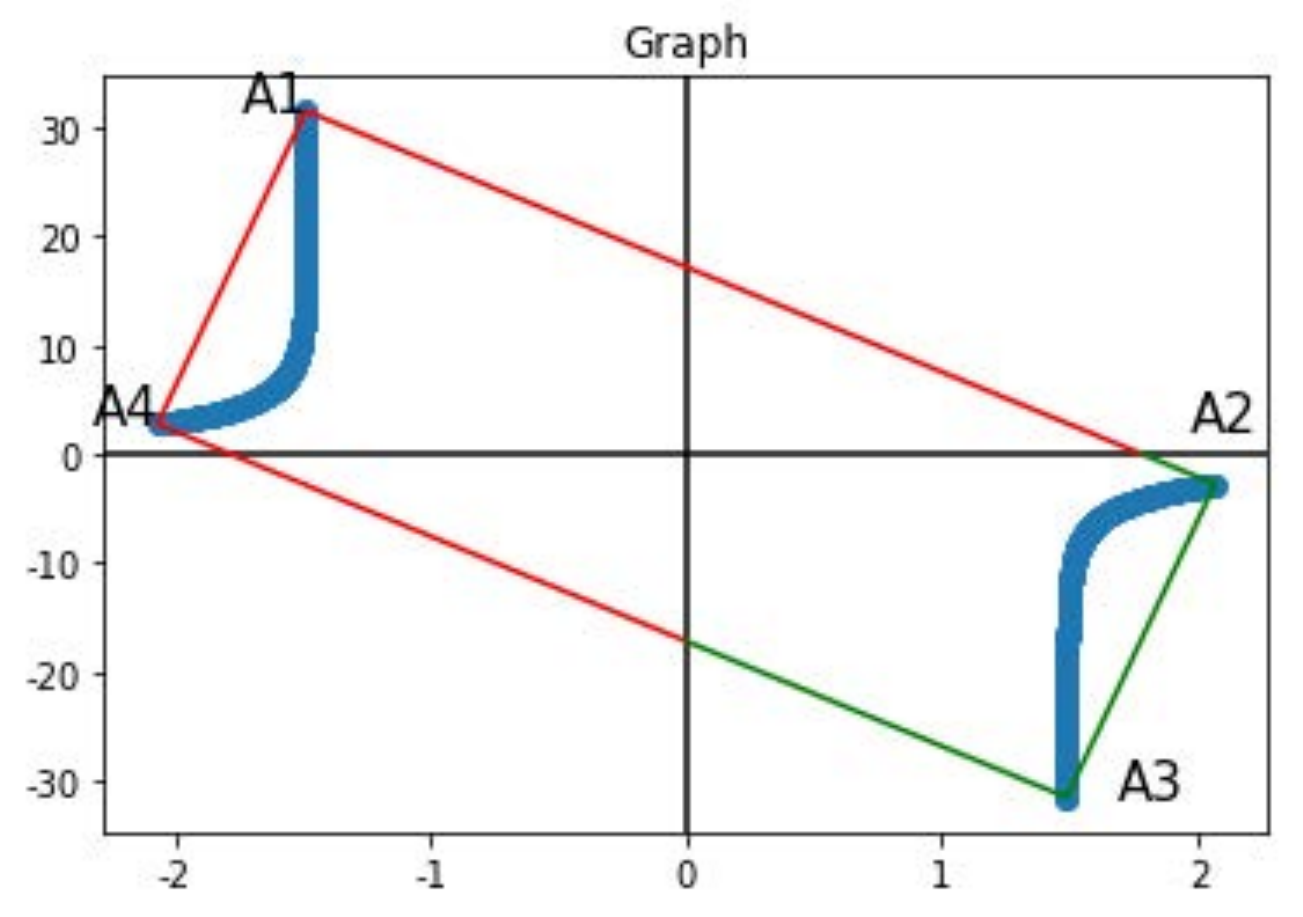}
	\caption{Convex hull based on the estimates from \cite{Amo2}. Green sides represent the possible crossing points of  $\nabla g(\boldsymbol{\theta})$.}\label{convex}
\end{figure}
The vector $\mathbf{c}$, defined 
 	as the gradient of $g(\boldsymbol{\theta})$ evaluated in the nominal values $(C_0,L_0)$, is given by
\begin{equation}
\mathbf c(\boldsymbol{\theta})^{T}=\displaystyle\frac{1}{2\sqrt{L_0}}\left(\displaystyle\frac{1}{C_0\sqrt{\log(C_0(T_0+1))}},\,
\displaystyle-\frac{\sqrt{\log(C_0(T_0+1))}}{L_0}\right)\label{grad};
\end{equation}
depending on the value of $T_0$, $\mathbf{c}$ has a different angle and the line directed by $\boldsymbol{c}$  crosses the convex hull in $\overline{A_1A_2}$ or $\overline{A_2A_3}$ (equivalently $\overline{A_3A_4}$ or $\overline{A_1A_4}$) {(see  Figure \ref{convex2})}.

The following proposition  provides the properties of the Elfving locus for any values of the  extremes of the experimental domain and for any possible choice of the nominal values  of the parameters, $(C_0,L_0)$.

\begin{proposition}\label{prop1}
Consider the curve \eqref{param2} and its reflection through the origin. Let $A_1, A_2, A_3, A_4$ be the outermost points: $A_1=(-x(b),-y(b))$, $A_2=(x(a), y(a))$, $A_3=(x(b),y(b))$, $A_4=(-x(a),-y(a))$.\\ Then,   for any value of $(C,L)$  such that $0<C<\exp(L\,\phi^2)$ and $L>0$,
   the convex hull of these curves is $\overline{A_1A_2A_3A_4}$.

\end{proposition}
\vskip.1truecm

 \begin{proof}
 
 The main point is to prove that the curve (\ref{param2}) is always above the segment $\overline{A_3A_2}$ and below the $\overline{A_1A_2}$. The reasoning will be organized in the following steps:
\begin{enumerate}
    \item Observe that $x(\phi)>0$ and $y(\phi)<0$ for any $\phi\in[a,\; b]$ since $L>0$ and $e^{\phi^2L}>C>0$ for any $\phi$. Then, the curve \eqref{param2} and the points $A_3,A_2$ are always in the fourth quadrant of the Cartesian plane, while its reflection and the points $A_4,A_1$ are always in the second quadrant.\\
    \item We have $x'(\phi)<0$ for any $\phi\in[a,\; b]$; it follows that $x(b)\le x(\phi)\le x(a)$.\\
    \item From the first equation of (\ref{param2}) we have $G(\phi,\,\boldsymbol{\theta})= Cx(\phi)$; moreover, by the definition of $G(\phi,\boldsymbol{\theta})$,
\begin{equation*}
\phi^2=\dfrac{1}{L}\log{\left(\dfrac{C^2x}{Cx-1}\right)};\nonumber
\end{equation*}
{\noindent then plugging into the second equation of (\ref{param2}), we obtain the cartesian equation of the curve:}
\begin{equation}\label{eq:cart}
y(x)=-\dfrac{C}{L}\, x\, \log{\left(\dfrac{C^2x}{Cx-1}\right)}, \quad x \in[x(b),x(a)].
\end{equation}
    \item Notice that $y\in {\cal C}^2([x(b),x(a)])$ and that
\begin{equation*}
y''(x)=-\dfrac{C}{Lx(Cx-1)^2}<0;
\end{equation*}
it follows that \eqref{eq:cart} is concave and therefore \eqref{param2} is above the segment $\overline{A_3A_2}$.\\
    \item In order to prove that \eqref{param2} is below the segment $\overline{A_1A_2}$ it is enough to prove that the tangent to the curve in $A_2$ is below $\overline{A_1A_2}$ (which has a negative slope $m$); this means that the slope of \eqref{eq:cart} in $x=x(a)$ is greater than the slope of $\overline{A_1A_2}$.\\
We have
\begin{equation}
y'(x)=\dfrac{C}{L}\left(\dfrac{1}{Cx-1}-\log{\dfrac{C^2x}{Cx-1}}\right)
\label{deriv1}
\end{equation}
and then
\begin{eqnarray}\label{der:xa}
y'(x)|_{x=x(a)}&=&\dfrac{C}{L}\left(\dfrac{1}{G(a,\boldsymbol{\theta})-1}-\log{\dfrac{C\, G(a,\boldsymbol{\theta})}{G(a,\boldsymbol{\theta})-1}}\right) \nonumber \\
&=& \dfrac{C}{L}\left( \dfrac{e^{a^2L}-C}{C}-a^2 L\right)\nonumber\\
&=& \dfrac{1}{L}e^{a^2L}-\dfrac{C}{L}-a^2C.
\end{eqnarray}
At this point there are two cases:\\
\begin{enumerate}
    \item[(a)] If $C<e^{a^2L}/(1+a^2L)$ then $y'(x)|_{x=x(a)}>0$ and it is straightforward that the slope of the curve is greater than the slope of $\overline{A_1A_2}$. Note that in this case we have $y'(x)>0$ for any $x\in[x(b),x(a)]$ (as in Figure \ref{convex}).\\
 \item[(b)] If $e^{a^2L}/(1+a^2L)<C<e^{\phi^2L}$, then $y'(x)|_{x=x(a)}<0$ (as in Figure \ref{convex3}) and we have to prove that
\begin{equation}\label{dis:slope}
y'(x)|_{x=x(a)}>m=-C\, \dfrac{a^2 G(a,\boldsymbol{\theta})+b^2 G(b,\boldsymbol{\theta})}{G(a,\boldsymbol{\theta})+G(b,\boldsymbol{\theta})}
\end{equation}
\end{enumerate}
Since \eqref{der:xa} can be written as
$$\dfrac{C}{L}\left( \dfrac{1}{G(a,\boldsymbol{\theta})-1}-a^2 L\right),$$
the inequality \eqref{dis:slope} is equivalent to
\begin{equation*}
\dfrac{1}{L} \dfrac{1}{G(a,\boldsymbol{\theta})-1}-a^2 >- \dfrac{a^2 G(a,\boldsymbol{\theta})+b^2 G(b,\boldsymbol{\theta})}{G(a,\boldsymbol{\theta})+G(b,\boldsymbol{\theta})},
\end{equation*}
 which gives
 \begin{equation*}
 \dfrac{1}{L} \dfrac{1}{G(a,\boldsymbol{\theta})-1}+ \dfrac{ (b^2-a^2) G(b,\boldsymbol{\theta})}{G(a,\boldsymbol{\theta})+G(b,\boldsymbol{\theta})}>0,
 \end{equation*}
 which is always satisfied since the left term is a sum of two positive quantities.
\end{enumerate}
\end{proof}


Denote by $(x_i,y_i)$, $i=1, ...,4$, the coordinates of the extremes $A_i$ of the convex hull stated in Proposition \ref{prop1} and denote by $\phi_i$ the corresponding values of $\phi$ in curve \eqref{param2} or in its symmetric (from Proposition 1, $\phi_i$ can be equal to $a$ or equal to $b$). Next proposition gives the $c$-optimal designs obtained by the crossing point between $\boldsymbol{c}=\nabla g(\boldsymbol{\theta})$ and the convex hull, according to the Elfving method, depending on the fixed value $T_0$.

\begin{proposition}\label{prop2}
	Depending on the fixed value of $T_0$, the convex hull  is crossed by  $\mathbf c(\boldsymbol{\theta})$  through  $\overline{A_iA_{i+1}}$, 	where $i=1, 2, 3$,  and  the $c$-optimal design is
		\begin{equation} \left\{
		\begin{array}{cc}
		\phi_i & \phi_{i+1} \\
		1-p_i & p_i \\
		\end{array}
		\right\},\qquad \mbox{ with  } p_i=\displaystyle\sqrt{\displaystyle\frac{(K x_0-y_i)^2+(x_0-x_i)^2}{(x_{i+1}-x_i)^2+(y_{i+1}-y_i)^2}}\,,
		\label{design}\end{equation}
		where $K=(\partial g/\partial L)/(\partial g/\partial C)$ and the coordinates of the crossing point $P_0$ are
		\begin{equation}
		x_0=\displaystyle\frac{y_i-\displaystyle\frac{y_{i+1}-y_i}{x_{i+1}-x_i}x_i}{K-\displaystyle\frac{y_{i+1}-y_i}{x_{i+1}-x_i}},\quad
		y_0=K x_0.\label{crossing}\end{equation}
	 In particular, let $$T_{0i}=\frac{1}{C}\exp{\left(-\frac{y_i L}{x_i C}\right)}-1,$$
		then
		\begin{itemize}
			\item for $T_0\in (max(0,\; {(1-C)}/{C}),\;T_{02}]$ the crossing point is in $\overline{A_1A_2}$;
			\item for $T_0\in (T_{02},\;T_{03}]$ the crossing point is in $\overline{A_2A_3}$;
			\item for $T_0>T_{03}$ the crossing point is  in $\overline{A_3A_4}$.
			\end{itemize}
		\end{proposition}

\vskip.1truecm
{\bf{Proof}} Observe that the lines that contain the segment $\overline{A_iA_{i+1}}$ and $\mathbf c(\boldsymbol{\theta})$ can be respectively written as
$$ y-y_i=\displaystyle\frac{y_{i+1}-y_i}{x_{i+1}-x_i}(x-x_i)\quad \mbox{and}\quad y=K x,$$
{\noindent hence the solution of the crossing point \eqref{crossing} follows straightforwardly.}

From the Elfving method we have that if the crossing point $P_0$ is in the side $\overline{A_iA_{i+1}}$, then the $c$-optimal design is given by \eqref{design} with $p_i=\|\overline{A_iP_0}\|/\|\overline{A_iA_{i+1}}\|$, where $\|\cdot\|$ is the euclidean norm.

Finally, taking into account that $c(\boldsymbol{\theta})$ is given by \eqref{grad};
{\noindent as $\log(C(T_0+1))>0$, then $$T_0>\displaystyle\frac{1-C}{C}.$$ Moreover, since  $\partial g/\partial L<0$ and $\partial g/\partial C>0$,
$\mathbf c(\boldsymbol{\theta})$ always moves into the fourth quadrant. As only the vertices $A_2$ and $A_3$ can be in the fourth quadrant, then  ${P_0}=A_i$, $i=2,\,3$,  are the  only two situations where the optimal design reduces to one point. In such a case,  ${y_i}/{x_i}=K$, and then }  $$T_{0i}=\displaystyle\frac{e^{\displaystyle-\frac{y_{i}L}{x_iC}}}{C}-1.$$
\bigskip

\begin{example}\label{ex1}
	Let ${\cal X}=[1.53,\,5.63]$, $C_0=0.671741$, $L_0=0.373098$ and $T_0=200$, then, according to Proposition \ref{prop2}, the convex hull is crossed by $\mathbf c(\boldsymbol{\theta})$ in $\overline{A_2A_3}$ and the optimal design is
	\begin{equation*}
	\xi^*_c= \left\{
	\begin{array}{cc}
	1.53 & 5.63\\
	0.5526 & 0.4474 \\
	\end{array}
	\right\}
	\end{equation*}
It is almost equally weighted as the D-optimal design.	Figure \ref{convex2} represents the convex hull and $\nabla g(\boldsymbol{\theta})$  in this example.
\end{example}

\begin{figure}[h]
	\centering
	\includegraphics[width=9cm]{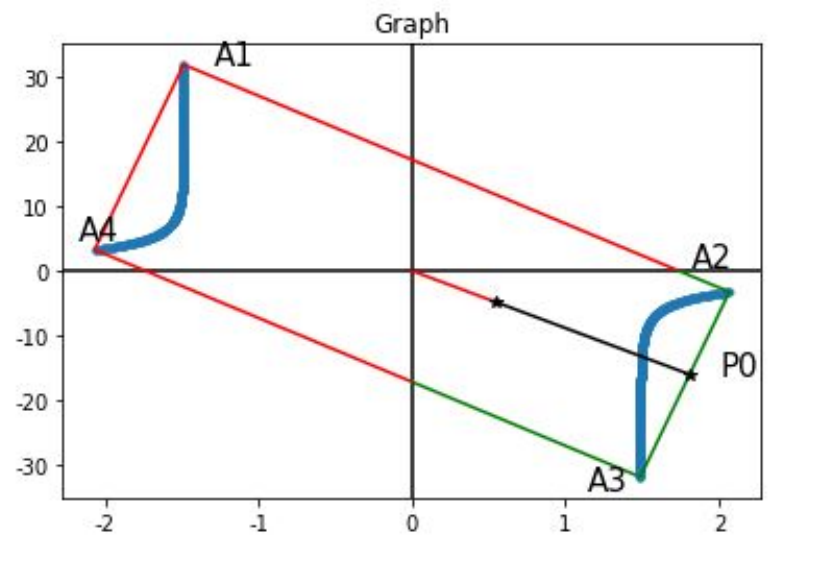}
	\caption{Location of the main points addressed in Proposition \ref{prop2} for the nominal values from \cite{Amo2} and $T_0=200$.}\label{convex2}
\end{figure}

\begin{figure}[h]
	\centering
	\includegraphics[width=9cm]{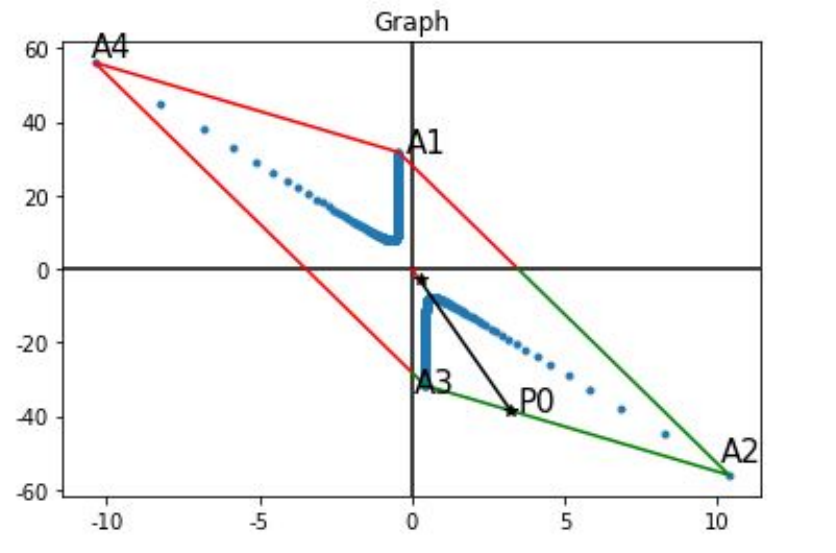}
	\caption{Convex Hull when Proposition \ref{prop1} (b) holds, for $T_0=2$, $C_0=2.3$ and $L_0$ and experimental domain as in \cite{Amo2}.}\label{convex3}
\end{figure}

\begin{example}\label{exb}
In the proof of  Proposition \ref{prop1} two situations are distinguished depending on whether point $A_2$ is a maximum or not.  One has been  illustrated in Example \ref{ex1} and the second one is illustrated in  this example. For  $L_0=0.373098$ and  $\phi\in[1.53,\,5.63]$, a value of  $C$ in the interval $(1.2784, 2.395)$ must be chosen, say $C_0=2.3$. Here $T_0=2$.

Figure \ref{convex3} represents the convex hull and $\nabla g(\boldsymbol{\theta})$ for $T_0=2$ in this example where $A_2$ is not a maximum. Now, Proposition 2 holds, and then
\begin{equation*}
	\xi^*_c= \left\{
	\begin{array}{cc}
	1.53 & 5.63\\
	0.2706 & 0.7294 \\
	\end{array}
	\right\}
	\end{equation*}
	
It is interesting to stress that this design put more weight in the right extreme, and therefore longer experimentation times are required,  although the limit $T_0$ is much smaller than in the previous example. 
\end{example}

\section{Sensitivity study}\label{sect:sensitivity}

Assume that $T_0$ is a given value; the following steps describe the procedure to perform a sensitivity study for the choice of the nominal values of the parameters.

\textbf {Step 1}:  Consider $\phi\in[1.53,\,5.63]$ and the nominal values $(C_0,L_0)$.  The c-optimal design is obtained  in Proposition \ref{prop2},
\begin{equation*}
\xi^{(0)}_c= \left\{
\begin{array}{cc}
\phi_{i}^{(0)}& \phi_{i+1}^{(0)}\\
1-p^{(0)}_i & p^{(0)}_i \\
\end{array}
\right\}
\end{equation*}

\textbf {Step 2}: We consider a grid where the parameters $C$ and $L$ take potential actual values $(C^*,L^*)$ in a neighborhood of the nominal values $(C_0,\,L_0)$.   Thus   we obtain the c-optimal design
\begin{equation*}
\xi^*_c= \left\{
\begin{array}{cc}
\phi_{i}^* & \phi_{i+1}^*\\
1-p_i^* & p_i^* \\
\end{array}
\right\},
\end{equation*}
when  the true values of the parameters is a pair $(C^*, L^*)$ in the grid.

\textbf {Step 3}: For each  $(C^{*},\, L^{*})$ in the grid, the following values are obtained:

$\qquad\bullet$ $$M_1=(1-p_i^*)I(\phi_{i}^*,C^{*},\,L^{*})+p_i^*I(\phi_{i+1}^*,C^{*},\,L^{*})$$
$$Var_1(g)=\nabla(g(C^{*},\,L^{*}))^T M_1^{-1} \nabla(g(C^{*},\,L^{*}))$$

$\qquad\bullet$ Consider the nominal values $(C_0,\, L_0)$ where $p^{(0)}_i$ and $\phi_{i}^{(0)}$ where obtained in Step 1 and obtain
$$M_0=(1-p^{(0)}_i)I(\phi_{i0},C^{*},\,L^{*})+p^{(0)}_iI(\phi_{i+1,0},C^{*},\,L^{*})$$
$$Var_0(g)=\nabla(g(C^{*},\,L^{*})^T M_0^{-1} \nabla g(C^{*},\,L^{*})$$

$\qquad\bullet$ Compute the relative efficiency given by $Var_1(g)/Var_0(g)$ .
\medskip

\begin{example}\label{ex2}

 From Proposition \ref{prop2},  three different situations can be distinguished depending on the nominal values chosen for $(C,\,L)$. In particular,
when  the nominal values $(C_0,\,L_0)$ are as  in the Example \ref{ex1}, $T_0$ can be  in the intervals (0.49, 2.57], (2.57, 203603.03] or (203603.03, $\infty$). From here, we consider the following  three cases:  $T_0 = 2$, $200$ and $300,000$. The first one is too small to have a practical interest, while the last one needs a diameter longer than those ones in the design space. Thus, they are extreme cases, but interesting to be considered in this study.

Consider a  grid of  points $(C^*,\,L^*)$ appropriate to detect sensitive changes in the efficiencies. In Figures  \ref{grid2}, \ref{grid200} and \ref{grid200000} the efficiencies for the three cases considered are shown.  In all the three cases $C^*$ varies in the interval $(C_0-0.3,\,C_0+0.3)$   while  $L^*$ varies  in the interval $(L_0-0.15, L_0+0.15)$ in  cases 1 and 2 and in the interval $(L_0-0.05, L_0+0.05)$ in case 3.

Observe that as we change $C^*$ and $L^*$ in the grid, also the three intervals stated in Proposition \ref{prop2}  change. Since the value $T_0$ is fixed, the crossing point can be in a different segment $\overline{A_iA_{i+1}}$ for $(C_0,\,L_0)$ and the point of the grid $(C^*,\,L^*)$. For instance, in Figures \ref{grid2} and \ref{grid200}, the largest decrement of the efficiency happens for large values of $C^*$ combined with small values of $L^*$, and it can be checked that they provide  values of $T_{02}$ smaller than 2 in the case 1 and $T_{03}$ values smaller or slightly larger than 200 in the case 2. Finally,  in Figure \ref{grid200000}, a smaller interval is chosen to vary  $L^*$ because dramatic changes of the efficiency are observed for further values of $L^*$; indeed, $T_{03}$ is also very sensitive to small changes in the parameters. Now, the efficiency decreases when $L^*$ grows and $C^*$ decreases (top left on Figure \ref{grid200000}) and when $L^*$ decreases and $C^*$ grows (bottom right on the table). These two situations correspond, respectively, with values of $T_{03}$ much smaller or much larger than 300,000. In other words, the cross points of the gradient with the convex hull are far away from the cross point of $(C_0,\,L_0)$ or in other segment.

We could say that, in the three cases, when both, $L^*$ and $C^*$, grow or decrease, the efficiency is more stable; but changes of $C^*$ and $L^*$ in opposite directions make the efficiency to reduce quicker.

\begin{figure}[H]
  \centering
\includegraphics[width=9cm]{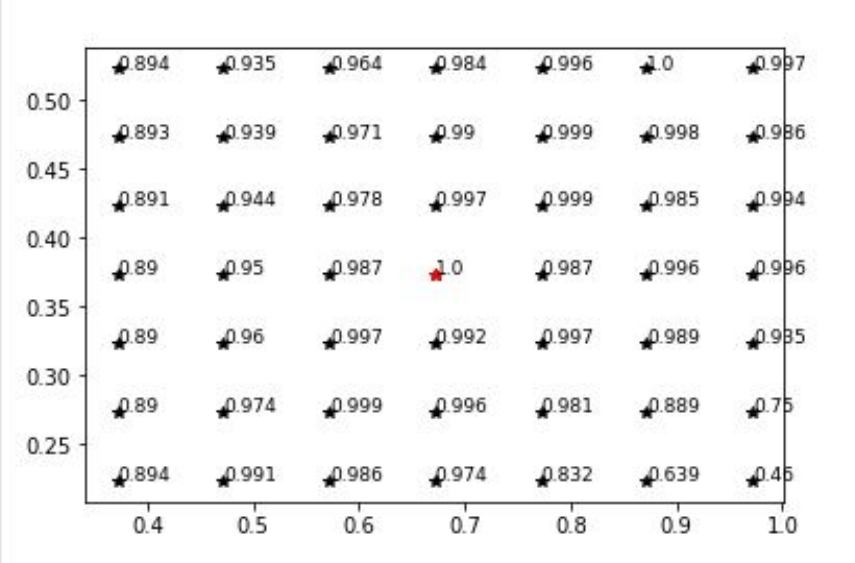}
  \caption{Efficiency values in each point of the grid for $T_0=2$.}\label{grid2}
\end{figure}

\begin{figure}[H]
  \centering
\includegraphics[width=9cm]{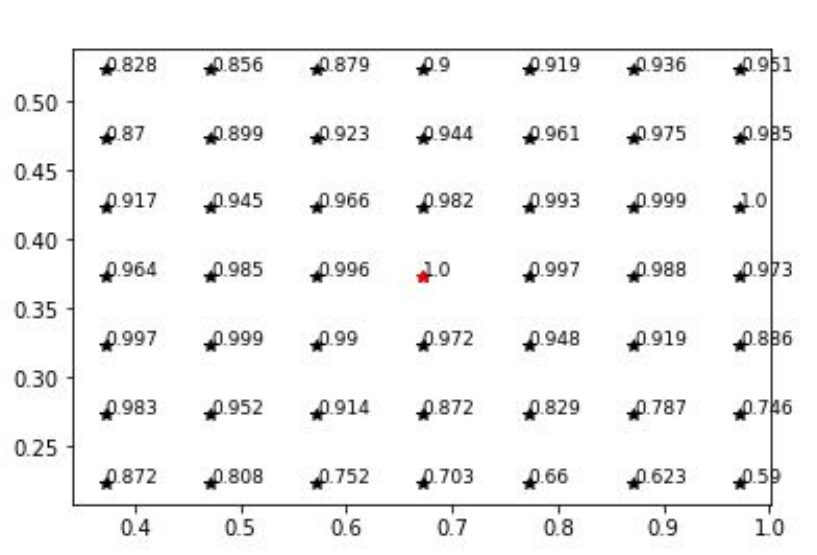}
  \caption{Efficiency values in each point of the grid for $T_0=200$.}\label{grid200}
\end{figure}

\begin{figure}[H]
  \centering
\includegraphics[width=9cm]{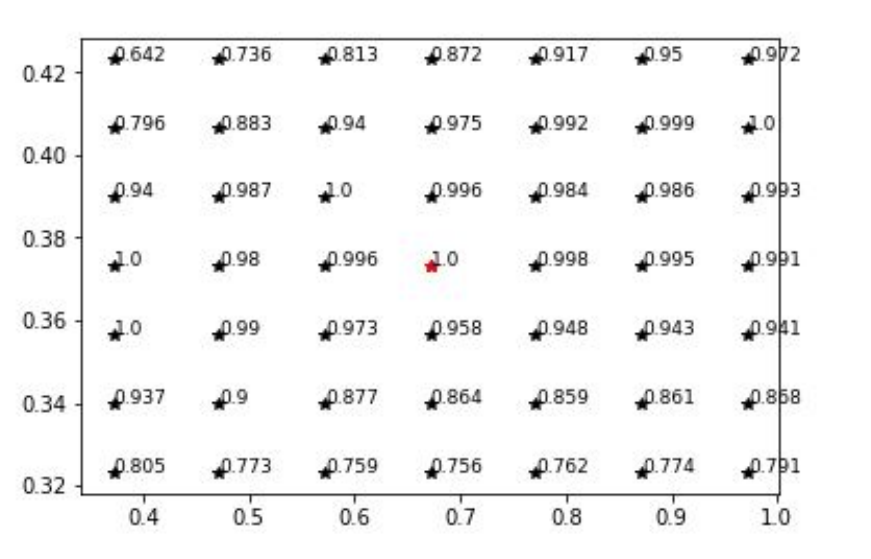}
  \caption{Efficiency values in each point of the grid for $T_0=300,000$.}\label{grid200000}
\end{figure}

\end{example}

\section{Consistency of the linearization procedure}

Observe that expressions (\ref{linearization}) and (\ref{c-var}) in Section \ref{sect:2} show the two linear approximation  procedures that have been adopted to solve the problem. The main goal of this section is to compare the a priori approximated variances and covariances of the estimates with the empirical variances and covariances of the estimate obtained by simulation. This is to have an insight of the accuracy of linearizing when looking for c-optimal designs in non-linear models.

The simulations will be performed in the following steps:

\textbf{Step 1}:  Consider  the nominal values $(C_0,\,L_0)$ and obtain the optimal design (\ref{design}) for a fixed value $T_0$.  Following the notation in Proposition \ref{prop2},  $n_i$ observations are randomly allocated at   $\phi_i$ and $n_{i+1}=n-n_i$ at $\phi_{i+1}$.

\textbf{Step 2}:   We obtain the MLEs of $C$, $L$ and  $g(\boldsymbol{\theta})$. As the responses follow an exponential distribution with mean (\ref{model1}), we have $n_i$ responses from an exponential distribution with parameter $\lambda_i$ that are denoted by $t_k^{(i)}$, $k=1,\cdots,n_i$ and $n_{i+1}$ with parameter $\lambda_{i+1}$, which are denoted by $t_k^{(i+1)}$, $k=1,\cdots,n_{i+1}$ where

\begin{equation} \lambda_j=\dfrac{C}{e^{L\phi_j^2}-C},\qquad j=i,\,i+1.\label{parameter}\end{equation}

The  likelihood function depends on the sample obtained, $\mathbf{t}$, and the parameter values $C$ and $L$, 

$$\mathcal{L}_n=\mathcal{L}_n(\mathbf{t},\,\theta)=\lambda_i^{n_i}e^{-\lambda_i\displaystyle\sum_{k=1}^{n_i}t_k^{(i)}}\lambda_{i+1}^{n_{i+1}}
e^{-\lambda_{i+1}\displaystyle\sum_{k=1}^{n_{i+1}}t_k^{(i+1)}}.$$

{\noindent By solving the equations $\partial\mathcal{L}_n/\partial C=0$ and  $\partial\mathcal{L}_n/\partial L=0$ we have that:}
$$\hat\lambda_j=\dfrac{n_j}{\displaystyle\sum_{k=1}^{n_j}t_k^{j}}=\dfrac{1}{\overline{T}_j};\qquad j=i,\,i+1$$
{\noindent Solving this system of equations we finally obtain the MLEs of $C$ and $L$:}
\begin{equation}
\hat{C}=\left(\dfrac{(1+\overline{T}_{i+1})^{\phi_i^2}}{(1+\overline{T}_i)^{\phi_{i+1}^2}}\right)^{\dfrac{1}{\phi_{i+1}^2-\phi_i^2}},\qquad
\hat{L}=\log\left(\dfrac{(1+\overline{T}_{i+1})}{(1+\overline{T}_i)}\right)^{\dfrac{1}{\phi_{i+1}^2-\phi_i^2}}
\end{equation}

The MLE of $g(\boldsymbol{\theta})$ in (\ref{gmodel1}) is given by $g(\hat{\boldsymbol{\theta}})$, where $\hat{\boldsymbol{\theta}}^T=(\hat{C},\, \hat{L})$.

\textbf{Step 3}: Step 2 is repeated  $m$ times obtaining three  $m$-vectors, $\mathbf{\hat{C}}$, $\mathbf{\hat{L}}$, $\mathbf{\hat{g}}$,  which contain, respectively, the MLEs of $C$, $L$ and $g(\boldsymbol{\theta})$ computed at each step.

\textbf{Step 4}: To study the accuracy of the approximation,  the covariance matrix of $\hat{\boldsymbol{\theta}}$ is approximated by  the empirical covariance matrix of $(\boldsymbol{\hat{C}},\;\boldsymbol{\hat{L}})$. Since the MLE is asymptotically efficient, the covariance matrix of $\hat{\boldsymbol{\theta}}$  should be similar to the Frechet-Cramer-Rao bound for $n$ sufficiently large.  In the multiparameter case, this bound is equal to $\Im=\partial\Psi/\partial \boldsymbol{\theta}^T \times I(\phi,\,\boldsymbol{\theta})*\partial\Psi^{T}/\partial\mathbf{\theta}$, where $I(\phi,\,\boldsymbol{\theta})$ is defined in \eqref{minf} and $\Psi(\boldsymbol{\theta})=E(\hat{\boldsymbol{\theta}}) $.

Observe that  $(\partial\Psi/\partial\boldsymbol{\theta})_{ij}=
\partial\Psi_i/\partial({\theta}_j)=Cov(\hat{\theta}_j,\;\partial log(\mathcal{L}_n)/\partial{\theta_j}))$, where $\mathcal{L}_n$ is the likelihood function.  In order to approximate this matrix, in step 2 we will also obtain, in each run, the bidimensional vector:
\begin{eqnarray}
\partial log(\mathcal{L}_n)/(\partial\boldsymbol{\theta})&=&\left(\begin{array}{c}
\partial log(\mathcal{L}_n)/\partial{C}\\
\partial log(\mathcal{L}_n)/\partial{L}
\end{array}\right)\nonumber\\
&=&\displaystyle\sum_{j=i}^{i+1}\displaystyle\sum_{k=1}^{n_j}
\left(\begin{array}{c}
\displaystyle\frac{1}{C_0}+
\displaystyle\frac{1}{e^{L_0\phi_j^2}-C_0}+
\displaystyle\frac{e^{L_0\phi_j^2}}{(e^{L_0\phi_j^2}-C_0)^2}t_k^{(j)}       \\
\displaystyle\frac{-\phi_j^2e^{L_0\phi_j^2}}{e^{L_0\phi_j^2}-C_0}
\left[1-\displaystyle\frac{C_0}{e^{L_0\phi_j^2}-C_0}t_k^{(j)} \right]
 \end{array}\right)\label{fn}
\end{eqnarray}
\noindent{then, we approximate  $(\partial\Psi/(\partial\boldsymbol{\theta}))_{ij}$ with the corresponding sample covariance.

\begin{example}\label{ex3}

Consider the setup of Example \ref{ex1}.

\textbf{Step 1:} Let   $C=0.671741$ and $L=0.373098$ as in \cite{Janda}, and consider several values of  $T_0$ (see Table \ref{simula}).

\textbf{Step 2:} We allocate randomly $n=1000$ experimental points following the optimal design obtained from Proposition \ref{prop2}.  The MLE values of $C$, $L$ and $g(\boldsymbol{\theta})$ are obtained jointly with the pair of values of the vector (\ref{fn}) that we denote, respectively, $\mathbf{f}_{n}^{(1)}$ and $\mathbf{f}_{n}^{(2)}$ .

\textbf{Step 3:} Step 2 is repeated $m=1000$ times and the 1000-dimensional vectors $\hat{C}$, $\hat{L}$, $g(\hat{\boldsymbol{\theta}})$, $\mathbf{f}_{n}^{1}$ and  $\mathbf{f}_{n}^{2}$ are stored.

\textbf{Step 4:} Table \ref{simula} shows a high similitude between the target value $g(\boldsymbol{\theta})$ and its MLE $\hat{g}$. Also, between the variance obtained with the simulated $Cov(\boldsymbol{\hat{C}},\;\boldsymbol{\hat{L}})$ denoted in the table as $\hat{Var}(\hat{g})$ and the variance obtained with $\Im$, which is denoted in the table as  $Var(\hat{g})$. The numbers must be multiplied by $10^{-4}$.

\begin{table}[!hbp]
\begin{tabular}{ccc|ccccc|ccc}
  \hline
   &\multicolumn{2}{c}{$T_0<T_{02}$}& \multicolumn{5}{c}{$T_{02}<T_0<T_{03}$}& \multicolumn{3}{c}{$T_0>T_{03}$}\\
  \hline
  $T_0$ &0.5& 2& 20& 200& 2000& $2\times10^4$& $2\times10^5$& $3\times10^5$& $6\times10^6$& $10^8$\\
  $g$ & 0.14 & 1.37 & 2.66 & 3.62 & 4.39 & 5.05 & 5.63 & 5.72&6.38& 6.95\\
  $\hat{g}$ & 0.23 & 1.37 & 2.66 & 3.62 & 4.39 & 5.05 & 5.63 & 5.72&6.38& 6.95\\
  p & 0.91 & 0.98& 0.21 & 0.45 & 0.66 & 0.84& 0.999 & 0.02 & 0.15 & 0.21 \\
  $\hat{Var}({g})^*$& 87 & 5.8 & 1.4 & 0.9 & 0.8 & 0.6 & 0.6 & 0.6 & 1.0 & 1.4 \\
  $Var(\hat{g})^*$& 562 & 6.1 & 1.4 & 0.9 & 0.8 & 0.6 & 0.6 & 0.6 & 1.0 & 1.4\\
  \hline
\multicolumn{7}{l}{ $^*$ The variances must be multiplied by $10^{-4}$} \\
  \caption{Simulation performance for Example \ref{ex1} for several values of $T_0$}\label{simula}
\end{tabular}
\end{table}

Observe that for $T_{0}=0.5$ nor the estimator, neither the variance are similar. As  $T_0$ is in the interval (0.4887, $\infty$), values close to the boundary carry out a slower convergence of the estimators. In Table \ref{convergence} we study the approach for $T_{0}=0.5$ of $g=0.1426$ and $\hat{g}$ and $Var[\hat{g}]$ and $ \hat{Var}[\hat{g}]$ for increasing values of the sample size $n$.
\begin{table}[!hbp]
\begin{tabular}{cccccc}
  \hline
  $n$ &1000& 5000& 10000& 100000& 1000000\\  \hline
 $bias=\hat{g}-g$&0.0874&0.0421&0.0218&0.0014&-0.0021\\
 $Var[\hat{g}]^*$&562.6&166.0&66.6&7.8&0.5\\
 $\hat{Var}[\hat{g}]^*$&87.3&44.8&25.5&7.6&0.6\\
   \hline
\multicolumn{6}{l}{ $^*$ The variances must be multiplied by $10^{-4}$} \\
\end{tabular}
\caption{Accuracy of the approximations for different values of $n$, $T_0=0.5$ and nominal values $C_0=0.671741$ and $L_0=0.373098$}\label{convergence}
\end{table}

The decreasing rate is smaller for the bias than it is for the variance.

\end{example}

\section{Conclusions}\label{sect:conclusions}

In this paper we consider the problem of estimating the parameters of a non-linear model  for the time between two jams  in the emptying of a silo. This may be applied to a number of phenomena such as delivering some material on a mine on a vertical tunnel. In most of the cases a jam might be rather dramatic involving some expense procedure to break the jam. In the case of the mine some explosive has to be use including risks and delays. Then a very important aim is to determine the diameter of the outlet, say $\phi$,  in order to guarantee a period of time long enough. This could be considered as a specific expected time, say $T_0$, or else a specific probability of reaching a specific time without jams.  This entails the estimation of a lower bound  expressed as a non-linear function that depends on the unknown parameters and $T_0$.  For both situations, expected time and probability,  give the same function of the parameters to be estimated tuning adequately the three specific constants mentioned above. In order to obtain an analytical solution of the problem, first we use the Fisher Information approximation for the covariance matrix of the estimates of the paramateres. Then the non-linear lower  bound, which is the target for estimation, is linearized being the its gradient the c-vector for c--optimality. A model with two parameters is chosen, and, so, the graphic Elfving procedure to find the c-optimal design is used.

 Propositions \ref{prop1} and \ref{prop2} establish, respectively, the main characteristics of the convex hull depending on the parameter values and then an explicit expression for the c-optimal design. Moreover, the latter indicates that the c-vector may intersect the convex hull in three sides of the convex hull depending on three intervals where  $T_0$ can lie. The vertices produce c-optimal designs with only one--point designs, otherwise two points are needed.

 The vertices of the convex hull are critical points in the sensitivity analysis since they indicate a change of the type of design. For this study a uniform grid with  values for the parameters around the nominal values was considered in order to detect big changes in the efficiency. A dramatic loss of efficiency happens when the parameter values considered in the grid produce a change of edge for the the crossing point  of the c-vector. A smaller decreasing is observed when the crossing point moves away on  the same edge of the convex hull. Both facts imply a very important change of the weights of the c-optimal design in Proposition \ref{prop2}. Besides this, for very  large values of $T_0$, the  sensitivity of the design with respect to the selection of the nominal values  is large, in fact, a small change of one of the parameters gives place to a dramatic decreasing of the efficiency, this is why the sensitivity study requires a reduced scale on this parameter. 
 
 A simulation study is carried out to check the accuracy of the double procedure to linearize the problem. So that, given the original non-linear model the ML estimators are obtained in a simulation procedure with a large number $n$ of observations allocated in the c-optimal design,  given a $T_0$ value and usual nominal values taken  from the  literature. Results show very close results, in general,  the approximation procedure produces slightly higher variances  of the lower bound for the silo outlet size than the simulated one. When $T_0$ is close to its lower bound, the convergence is slower and $n$ must be enlarged.

\section*{Suppementary material} All the computations have been done with Python 3.7. Codes are provided in two files.

\section*{Acknowledgements}
The first author was  sponsored by Ministerio de Ciencia y Tecnolog\'{\i}a MTM2016-80539-C2-1-R and the third one by Ministerio de Ciencia y Tecnolog\'{\i}a MTM2017-83812-P and MTM2016-77015-R. The second author thanks the Departamento de Estad\'{\i}stica, Inform\'atica y Matem\'aticas  of the Universidad P\'ublica de Navarra  for enduring his scientific visit to the department.

\end{document}